\documentclass[acus]{JAC2000}
\usepackage{epsfig,amssymb}    

\def\3{\ss}
\newcommand{\be}{\begin{equation}}
\newcommand{\ee}{\end{equation}}
\newcommand{\bea}{\begin{eqnarray}}
\newcommand{\eea}{\end{eqnarray}}

\usepackage{graphicx}

\begin{document}
\title{
IMPEDANCE OF A BEAM TUBE WITH SMALL CORRUGATIONS
\protect\thanks{Work supported by the U.S.\ Department
of Energy under contract DE-AC03-76SF00515.
}\vspace*{-4mm}}

\author{K.L.F.\ Bane, G. Stupakov, SLAC, Stanford University,
 Stanford, CA 94309, U.S.A.}
\maketitle

\vspace*{-2.3cm}
\section{Introduction}\label{sec:intro}
In accelerators with very short bunches, such as is envisioned in
the undulator region of the
Linac Coherent Light Source (LCLS)\cite{LCLS}, the
wakefield due to the roughness of the beam-tube walls
can have important implications on the required smoothness and
minimum radius allowed for the beam tube.
Of two theories of roughness impedance, one yields
an almost purely inductive impedance\cite{BCN}, the other a single resonator
impedance\cite{MN};
for smooth bunches, whose length is large compared to the wall perturbation
size, these two models give comparable results\cite{BN}.

Using very detailed, time-domain simulations
it was found in Ref.~\cite{MN}
that a beam tube with a random, rough surface has
an impedance that is similar to that of
one with small, {\it periodic} corrugations.
It was further found that the wake was similar to that
of a thin dielectric layer (with dielectric constant $\epsilon\approx2$)
on a metallic tube: $W_z(s)\approx2{\cal K}_0\cos k_0s$, with
wave number and loss factor
\begin{equation}
k_0={2\over \sqrt{a\delta}}\quad{\rm and}\quad
{\cal K}_0={Z_0c\over2\pi a^2}
 \ ;
\label{resz}
\end{equation}
with $a$ the tube radius, $\delta$
depth of corrugation,
and $Z_0=377$~$\Omega$.
For the periodic corrugation problem this result was inferred
from simulations for which the period
$p\sim\delta$.
On the other hand, at the extreme of a tube with shallow oscillations, 
with $p\gg\delta$, 
the impedance was found, by a perturbation calculation of Papiernik,
to be composed of many weak, closely spaced modes
beginning just above pi phase advance\cite{papiernik79c}.

In this report we
find the impedance for two geometries of periodic, shallow corrugations:
one, with rectangular corrugations using a field matching approach,
the other, with 
smoothly varying oscillations
using a more classical perturbation approach.
In addition, we explore how these results change character
as the period-to-depth of
the wall undulation increases, and then
compare the results of the two methods.

\section{Rectangular Corrugations}

\begin{figure}[t]
\vspace*{-1.25cm}
\centering
\epsfig{file=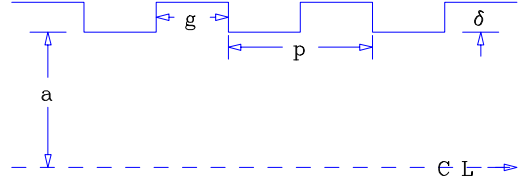, width=67.5mm}
\caption{The geometry considered.}
\label{figeom}
\end{figure}

Let us consider a 
cylindrically-symmetric
beam tube with the geometry shown in Fig.~\ref{figeom}.
We limit consideration here to the case $\delta/a$ small; for the moment,
in addition, let $\delta/p\gtrsim1$.
We follow the formalism of the field matching program TRANSVRS\cite{TRANS}:
In the two regions, $r\leq a$ (the tube region, Region I)
and $r\geq a$ (the cavity region, Region II) the Hertz vectors
are expanded in a complete, orthogonal set; $E_z$ and $H_\phi$ are matched
at $r=a$; using orthogonality properties an infinite dimensional,
 homogeneous matrix 
equation is
generated; this matrix is truncated; and finally,
the eigenfrequencies are found by setting its determinant
 to zero.
We demonstrate below that, for our parameter regime,
the system matrix can be reduced to dimension 1, and the results become
quite simple.

In the tube region, the $z$-component of the Hertz vector
\begin{equation}
\Pi_z^{I}= -\sum_{n=-\infty}^\infty {A_n\over\chi_n^2}
  {I_0(\chi_n r)\over I_0(\chi_n a)}e^{-j\beta_n z}\ ,
\end{equation}
with $I_0$ the modified Bessel function of the first kind, and
\begin{equation}
\beta_n= \beta_0+{2\pi n\over p}\ ,\quad\quad
\chi_n^2=\beta_n^2-k^2\ ,
\end{equation}
with $\beta_0$ the phase advance and $k$ the wave number of the mode.
In the cavity region,
\begin{equation}
\Pi_z^{II}= -\sum_{s=0}^\infty { C_s\over\Gamma_s^2}
  {R_0(\Gamma_s r)\over R_0(\Gamma_s a) }\cos[\alpha_s(z+g/2)]\ ,
\end{equation}
\begin{equation}
\alpha_s={\pi s\over g}\ ,\quad\quad
\Gamma_s^2= \alpha_s^2-k^2\ ,
\end{equation}
\begin{equation}
R_0(\Gamma_s r)= K_0(\Gamma_s [a+\delta])I_0(\Gamma_s r)-
                 I_0(\Gamma_s [a+\delta])K_0(\Gamma_s r)\ ,
\end{equation}
with $K_0$ the modified Bessel Function of the second kind.

$E_z$ and $H_\phi$ are given by
\begin{equation}
E_z= \left({\partial^2\over \partial z^2} +k^2\right)\Pi_z\ ,\quad\quad
Z_0H_\phi= -jk{\partial\Pi_z\over \partial r}\ .
\end{equation}
Matching these fields at $r=a$, and using the orthogonality of
$e^{-\beta_n z}$ on $[-p/2,p/2]$, and $\cos[\alpha_s(z+g/2)]$
on $[-g/2,g/2]$ we obtain a homogeneous matrix equation.
To find the frequencies, the determinant is set to zero; {\it i.e.}
\begin{equation}
{\rm det}\left[{\cal R}-\left({2g\over p}\right)N^T{\cal I}N\right]
=0\ ,\label{eqdet}
\end{equation}
with the matrix $N$ given by
\begin{equation}
N_{ns}= {2\beta_n\over(\beta_n^2-\alpha_s^2)g}
  \left\{ \begin{array}
          {r@{\quad:\quad}l}
   \sin (\beta_n g/2) & s\ {\rm even} \\
   \cos (\beta_n g/2) & s\ {\rm odd}\end{array}\right.\ ,
\end{equation}
and the diagonal matrices ${\cal R}$ and ${\cal I}$ by
\begin{equation}
{\cal R}_s= (1+\delta_{s0})ka\left(R_0^\prime\over xR_0\right)_{\Gamma_s a}
\ ,\quad\quad
{\cal I}_n= ka\left(I_0^\prime\over xI_0\right)_{\chi_n a}\ .
\end{equation}

For the beam, on average,
 to interact with a mode, one space harmonic
of the mode must be synchronous.
We will pick the $n=0$ space harmonic to be the synchronous one;
{\it i.e.} let $\beta_0=k$ (we take the particle velocity to be $v=c$).
We truncate the system matrix to dimension 1, keeping only the $n=0$ and
$s=0$ terms in the calculation (the other terms are small).
Now if $k\delta$ is small, then the $s=0$ term in ${\cal R}$ becomes
${\cal R}_0=2/(k\delta)$,
the $n=0$ term in ${\cal I}$ is ${\cal I}_0= ka/2$, and
$N_{00}\approx1$.
Eq.~\ref{eqdet} then yields 
\begin{equation}
k= \sqrt{2p\over a\delta g}\ ,
\label{eqktwo}
\end{equation}
which, for $p=2g$, equals $k_0$ of Eq.~\ref{resz}.

The loss factor is given by ${\cal K}=|V|^2/[4Up(1-\beta_g)]$\cite{argonne},
 with
$V$ the voltage lost by the beam to the mode, $U$
the energy stored in the mode, and $\beta_g$ the group velocity
over $c$.
The voltage lost in one cell is given by the synchronous $(n=0)$
space harmonic: $V=A_0p$, and the energy stored in one cell,
$U= 1/(2Z_0c)\int E\cdot E^*\,dv$,
is approximately that which is in the $n=0$ space harmonic:
$U=\pi A_0^2a^2p(1+k^2a^2/8)/(2Z_0c)$
 (for details, see Ref.~\cite{TRANS}).
For $\beta_g$, we take
 Eq.~\ref{eqdet} truncated to dimension 1, and 
expand near the synchronous point.
Taking the derivative with respect to $\beta_0$
and then setting $\beta_0=k$
we obtain:
\begin{equation}
(1-\beta_g)= {4\delta g\over ap}\ .
\end{equation}
 The loss factor becomes ${\cal K}={\cal K}_0$. 

The above method can be extended to modes of higher multipole
 moment $m$, in which case the beam will excite hybrid modes
rather than the pure TM modes of above\cite{TRANS}.
Again the system matrix can be reduced to the $n=0$ and $s=0$ terms,
and the lowest mode wave
number and loss factor have a simple form (for $1\leq m\ll a/\delta$):
\begin{equation}
k= \sqrt{(m+1)p\over a\delta g}\quad{\rm and}\quad
{\cal K}={Z_0c\over\pi a^{2(m+1)}}\ ,
\end{equation}
and $(1-\beta_g)= m(m+2)\delta g/(ap)$.
We note that the dipole $(m=1)$ frequency is
equal to the monopole $(m=0)$ frequency. Also, the 
wake at the origin is the same as for the resistive-wall wake
of a cylindrical tube\cite{chao},
as we expect.

\begin{figure}[htb]
\centering
\epsfig{file=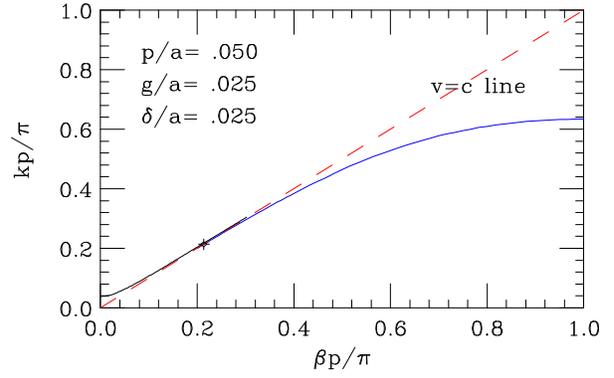, height=49.mm}
\caption{Dispersion curve example.}
\label{fidisp}
\end{figure}
\begin{figure}[htb]
\centering
\epsfig{file=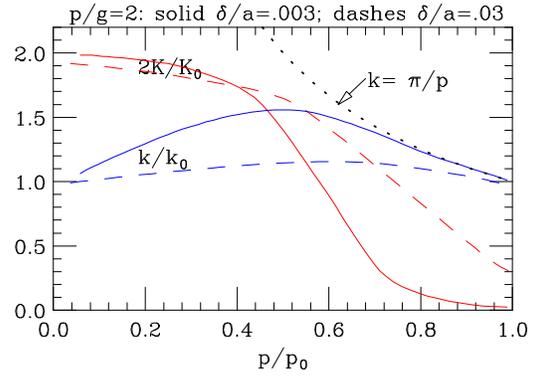, height=49.mm}
\caption{An example showing the effect of varying $p$.}
\label{fivarp}
\end{figure}

Running TRANSVRS with a matrix of dimension 40, we obtain 
a typical dispersion curve (see Fig.~\ref{fidisp}).
Here $k/k_0=1.07$, ${\cal K}/{\cal K}_0=.94$.
Note that even when $\delta/a$ is not so small, {\it e.g.}
for bellows with $\delta/a\approx.2$\cite{RRuth},
 the analytical formulas are still useful.
 Fig.~\ref{fivarp} shows how the strength and
frequency of the mode change as the period of undulation
is increased. The scale over which ${\cal K}$ drops to zero is
$p_0\approx\pi\sqrt{a\delta g/2p}$. 
By $p\sim p_0$, the one dominant mode has disappeared,
and we are left with the many
weak, closely spaced modes, beginning 
just above $kp=\pi$, that were found by
 Papiernik.

\section{Sinusoidal Corrugations}
Let us assume now that the pipe surface is given by
    \be
    r=a-h\sin \kappa z,
    \ee
where $2\pi/\kappa$ is the period of corrugation, and $h$ is its
amplitude. We assume that both the amplitude and the wavelength are
small, $h \ll a$ and $\kappa a \gg 1$. This allows us to neglect the
curvature effects and to consider the surface locally as a plane one.
We will also assume a shallow corrugation
    $h\kappa \ll 1$,
{\it i.e.} the amplitude of oscillation is much smaller than the period.

Introducing a local Cartesian coordinate system $x$, $y$, $z$ with
$y=a-r$ (directed from the wall toward the beam axis), and $x$
directed along $\phi$, the surface equation becomes $y = y_0(z)
\equiv h\sin \kappa z$. The magnetic field near the surface $H_x (y,z)$
does not depend on $x$ (that is $\phi$) due to the axisymmetry of
the problem. It satisfies the Helmholtz equation
    \be \label {helm}
    \frac{\partial^2 H_x}{\partial y^2}
    +
    \frac{\partial^2 H_x}{\partial z^2}
    +
    k^2 H_x
    =0
    \ee
with the boundary condition
    \be \label{boundary_cond2}
    (\vec{n}\nabla H)|_{y=y_0} =0,
    \ee
where $\vec{n}$ is the normal vector to the surface, $\vec{n} =
(0,1,-h\kappa \cos \kappa z)$.

Note that the longitudinal electric field $E_z$ can be expressed in terms of
$H_x$,
    \be \label{Ez1}
    E_z=-\frac{i}{k}
    \frac{\partial H_x}{\partial y}.
    \ee
Using the small parameter $h/a$, we will develop a perturbation theory for
calculation of $H_x$ near the surface and find how $E_z$ is related
to $H_x$.

In the zeroth approximation, the $z$ dependence of $H_x$ is dictated
by the beam current periodicity,
    \be \label{Hx}
    H_x(y,z) = {\cal H}(y)e^{ikz}.
    \ee
Putting Eq. (\ref{Hx}) into Eq. (\ref{helm}) we find that $d^2 {\cal
H}/dy^2 = 0$, hence ${\cal H}(y) = H_0+Ay$, where the  constant $A$
can be related, through Eq. (\ref{Ez1}), to the electric field on the
surface, $A=ikE_z$. We will see below that $A$ is second order
in $h$.

For a flat surface, for which $\vec{n} = (0,1,0)$, from the boundary
condition (\ref{boundary_cond2}), we would conclude that $A=0$,
however, the corrugations result in a nonzero $A$, and hence $E_z$.
Substituting the magnetic field (\ref{Hx}) into the right hand side
of Eq.  (\ref{boundary_cond2}) one finds
    \[
    \vec{n}\nabla H =
    -\frac{1}{2}ihk\kappa H_0
    \left[e^{i(k+\kappa) z}-e^{i(k-\kappa) z}\right]
    -ik\zeta H_0 e^{ikx}.
    \]
Clearly, the boundary condition is not satisfied in this
approximation. To correct this, we have to add satellite modes to the
fundamental solution (\ref{Hx})
    \be
    H_x(y,z) = {\cal H}(y)e^{ikz}+ {\cal H}_1(y,z),
    \ee
    \be
    {\cal H}_1(y,z) = B^+(y)e^{i(k+\kappa) z}
    +
    B^-(y)e^{i(k-\kappa) z}.
    \ee
The dependence of $B^{\pm}$ versus $y$ can be found from the
Helmholtz equation,
    \be \label{satellite_harm}
    B= B_0^{\pm}e^{-y\sqrt{\kappa^2 \pm 2\kappa k} },
    \ee
where $B_0^{\pm}$ are constants.
In order for $B^{\pm}$ to exponentially decay in $y$, we have to
assume here that $k < \kappa/2$.

Substituting ${\cal H}_1$ terms into the boundary condition
(\ref{boundary_cond2}) generates first order terms that have
$x$-dependence $\exp i(k\pm \kappa)x$, and second order terms
proportional to $\exp(i k x)$. From the former one finds that
    \be
    B_0^{\pm} =
    -\frac{ik\kappa H_0 h}{2\sqrt{\kappa^2 \pm 2k\kappa}},
    \ee
and the latter gives an expression for the tangential electric field
on the surface,
    \be \label{zeta_corrug_wall}
    E_z = \frac{1}{4}i k h^2 \kappa H_x
    \frac{\sqrt{\kappa^2 + 2k\kappa}+\sqrt{\kappa^2 - 2k\kappa}}
    {\sqrt{\kappa^2 - 4k^2}}.
    \ee

One can now solve Maxwell's equations with the boundary condition
given by Eq. (\ref{zeta_corrug_wall}) (see details in
\cite{stupakov99}). It turns out, that in the region of frequencies
$k < \kappa/2$ there exists a solution corresponding to a wave
propagating with the phase frequency equal to the speed of light.
\begin{figure}[htb]
\centering
\epsfig{file=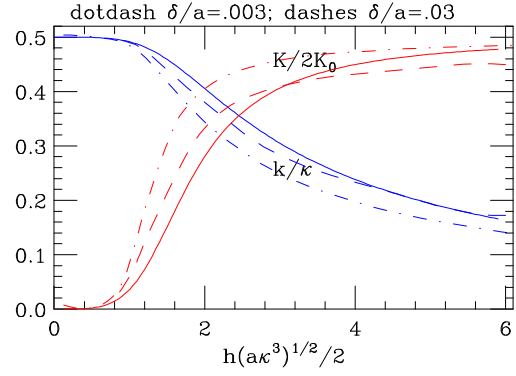, height=49.mm}
        \caption
        {
        Frequency and loss factor as function of height.
        \label{optical}
        }
        \end{figure}
The frequency and the loss factor of the mode are shown in
Fig.~\ref{optical} (solid lines). We see that  decreasing the height of the
corrugation results in smaller wakes, and hence leads to the
suppression of the interaction of the synchronous wave with the beam.
In the limit of small frequencies, $k \ll \kappa$ the frequency is
    \be
    k_1 = \frac{2}{h\sqrt{a\kappa}}\,.
    \ee

We should mention here that the perturbation theory breaks down for
very small values of $h$. Indeed, we implicitly assumed that the
satellite harmonics in Eq. (\ref{satellite_harm}) are localized near
the surface, otherwise our approximation of plane surface becomes
invalid. Hence, we have to require that $\kappa - 2k \gg a^{-1}$,
which gives the following condition of applicability:
$    h >  a^{-1/4}\kappa^{-5/4}$.
This condition explains why this mode was not found by Papiernik:
 being perturbative in parameter $h$ the approach
developed in his paper is applicable only when $h$ can be made
arbitrarily small.
Let us also mention
 that our approach can be applied to the rectangular
corrugations problem: we take Faraday's law, to obtain the averaged
longitudinal field at $r=a$, $\langle E_z\rangle=ik\delta gH_x/p$, and then
follow the method of Ref.~\cite{stupakov99} to reproduce
 Eq.~\ref{eqktwo}.

Finally, in Fig.~\ref{optical} we include also the results
of Fig.~\ref{fivarp}, obtained by field matching (the dashes, the
dotdashes).
For the comparison we make the correspondences
with the first Fourier component of the wall shape: $\kappa=2\pi/p$ and
$h=2\delta/\pi$. We note that even though the rectangular corrugations 
violate our requirement for smoothness, the results
for the two methods are very similar.

Acknowledgement:
We thank A. Novokhatskii for his contribution to our
understanding of the problem of roughness impedance.


\vspace*{-2mm}
\begin{thebibliography}{9}
\small
\vspace*{-1.0mm}
\bibitem{LCLS}
Linac Coherent Light Source (LCLS) Design Study Report. SLAC-R-521,
Apr 1998. 
\vspace{-1.5mm}
\bibitem{BCN}
K. Bane, {\it et al}, PAC97, p. 1738 (1997);
G.V. Stupakov, {\it Phys. Rev. AB} {\bf 1}, 64401 (1998).
\vspace{-1.5mm}
\bibitem{MN}
A. Mosnier and A. Novokhatskii, PAC97, p. 1661 (1997).
\vspace{-1.5mm}
\bibitem{BN}
K. Bane and A. Novokhatskii, SLAC-AP-177, March 1999.
\vspace{-1.5mm}
\bibitem{papiernik79c}
{M.~Chatard-Moulin} and {A.~Papiernik},
{\it IEEE Trans. Nucl. Sci.} {\textbf{26}},
{3523} ({1979}).
\vspace{-1.5mm}
\bibitem{TRANS}
  K. Bane and B. Zotter, Proceedings of the 11$^{\rm th}$ Int. Conf.
  on High Energy Accelerators, CERN, p.~581 (1980).
\vspace{-1.5mm}
\bibitem{argonne}
  See, {\it e.g.}, E. Chojnacki, {\it et al}, PAC93, p. 815, 1993;
A.~Millich, L.~Thorndahl, CERN-CLIC-NOTE-366, Jan. 1999.
\vspace{-1.5mm}
\bibitem{chao}
    A.\ Chao, ``Physics of Collective Instabilities
    in High-Energy\vspace{-.7mm} Accelerators'', John Wiley \& Sons, New York
    (1993).
\vspace{-1.5mm}
\bibitem{RRuth}
K. Bane and R. Ruth, SLAC-PUB-3862, January 1986.
\vspace{-1.5mm}
\bibitem{stupakov99}
G.V. Stupakov in {T. Roser and S. Y. Zhang}, eds.,
 {AIP} Conference Proceedings 496, 1999, p. 341.
\end{thebibliography}
\end{document}